\documentclass[runningheads,psfig]{svmult}
\usepackage{psfig}
\usepackage{makeidx}   % allows index generation
\usepackage{graphicx}  % standard LaTeX graphics tool
\usepackage{subeqnar}  % subnumbers individual equations
\usepackage{multicol}  % used for the two-column index
\usepackage{cropmark} % cropmarks for pages without
\usepackage{physprbb}  % centered layout of diverse elements, etc.
\usepackage{amsmath}
\usepackage{amssymb}

\def\lcdm{$\Lambda$CDM}

\def\gsim{${\raise.3ex\hbox{$>$\kern-.75em\lower1ex\hbox{$\sim$}}}$}
\def\lsim{${\raise.3ex\hbox{$<$\kern-.75em\lower1ex\hbox{$\sim$}}}$}

\begin{document}
\title*{Cosmological Parameters}
\toctitle{Cosmological Parameters}
% allows explicit linebreak for the table of content
%
%
\titlerunning{Cosmological Parameters}
% allows abbreviation of title, if the full title is too long
% to fit in the running head
%
\author{Joel R. Primack}
\authorrunning{Primack}
% if there are more than two authors,
% please abbreviate author list for running head
%
%
\institute{Physics Department, University of California, Santa Cruz,
CA 95064 USA}

\maketitle              % typesets the title of the contribution

\begin{abstract}
The cosmological parameters that I will emphasize are the Hubble
parameter $H_0 \equiv 100 h$ km s$^{-1}$ Mpc$^{-1}$, the age of the
universe $t_0$, the average matter density $\Omega_m$, the baryonic
matter density $\Omega_b$, the neutrino density $\Omega_\nu$, and the
cosmological constant $\Omega_\Lambda$. The evidence currently favors
$t_0 \approx 13$ Gyr, $h \approx 0.65$, $\Omega_m \approx 0.4\pm0.1$,
$\Omega_b = 0.02h^{-2}$, $0.001 < \Omega_\nu < 0.1$, and
$\Omega_\Lambda \approx 0.7$.
\end{abstract}

\section{Introduction}
In this review I will concentrate on the values of the cosmological
parameters. The other key questions in cosmology today concern the
nature of the dark matter and dark energy, the origin and nature of
the primordial inhomogeneities, and the formation and evolution of
galaxies.  I have been telling my theoretical cosmology students for
several years that these latter topics are their main subjects for
research, since determining the values of the cosmological parameters
is now mainly in the hands of the observers.  

In discussing cosmological parameters, it will be useful to
distinguish between two sets of assumptions: (a) general relativity
plus the assumption that the universe is homogeneous and isotropic on
large scales (Friedmann-Robertson-Walker framework), or (b) the FRW
framework plus the \lcdm\ family of models.  In addition to the FRW
framework, the \lcdm\ models assume that the present matter density
$\Omega_m$ plus the cosmological constant (or its equivalent in ``dark
energy'') in units of critical density $\Omega_\Lambda = \Lambda/(3
H_0^2)$ sum to unity ($\Omega_m + \Omega_\Lambda = 1$) to produce the
flat universe predicted by simple cosmic inflation models.  These
\lcdm\ models assume that the primordial fluctuations were adiabatic
(all components fluctuate together) and Gaussian, and had a Zel'dovich
spectrum ($P_p(k)=Ak^n$, with $n \approx 1$), and that the dark matter
is mostly of the cold variety.  

Although the results from the Long-Duration BOOMERANG
\cite{boom,boomparam} and the MAXIMA-1 \cite{maxima,maximaparam} CMB
observations and analyses \cite{jaffe} were were not yet available at
the Dark Matter 2000 conference, I have made use of them in preparing
this review.  The table below summarizes the current observational
information about the cosmological parameters, with estimated
$1\sigma$ errors.  The quantities in brackets have been deduced using
at least some of the \lcdm\ assumptions.  The rest of this paper
discusses these issues in more detail.  But it should already be
apparent that there is impressive agreement between the values of the
parameters determined by various methods.

\begin{table}
\caption{Cosmological Parameters [results assuming \lcdm\ in
brackets]}
\label{ta:parameters}
%\centerline{\vbox{\halign{\ \ #\hfill \quad \qquad &$#$\hfill \ &$#$
%&#\hfill \ \ \cr
\centerline{\vbox{\halign{\ $#$\hfill \ &$#$ &#\hfill \ \ \cr
\noalign{\hrule}
\noalign{\vskip .10in}
H_0          &= &$100 \,h$ km s$^{-1}$ Mpc$^{-1}$ ,
                                \ $h = 0.65 \pm 0.08$ \cr
t_0          &= &9-16 Gyr (from globular clusters) \cr
{}           &= &[9-17 Gyr from expansion age, \lcdm\ models] \cr
\Omega_b     &= &$(0.045\pm0.0057)h_{65}^{-2}$ (from D/H) \cr
{}           &> &[$0.04 h_{65}^{-2}$ from Ly$\alpha$ forest opacity] \cr
\Omega_m     &= &$0.4\pm0.2$ (from cluster baryons) \cr
{}           &= & [$0.34\pm0.1$ from Ly$\alpha$ forest $P(k)$] \cr
{}           &= & [$0.4\pm0.2$ from cluster evolution] \cr
{}           &> & $0.3$ ($2.4 \sigma$, from cosmic flows) \cr
{}           &= & [$0.5 \pm 0.1$ from flows plus SN Ia]\cr
{}           &\approx &$\frac{3}{4} \Omega_\Lambda - \frac{1}{4}
                        \pm{\frac{1}{8}}$ from SN Ia \cr
\Omega_m + \Omega_\Lambda  &= &$1.11\pm0.07$ (from CMB peak location) \cr
\Omega_\Lambda &= &$0.71 \pm 0.14$ (from previous two lines) \cr
{}           &< & 0.73 (2$\sigma$) from radio QSO lensing \cr
\Omega_\nu  &$\gsim$ &0.001 (from Superkamiokande) \cr
{}           &$\lsim$ &[0.1] \cr
\noalign{\vskip .10in}
\noalign{\hrule}
}}}
\end{table}

\section{Age of the Universe $t_0$}

The strongest lower limits for $t_0$ come from studies of the stellar
populations of globular clusters (GCs).  In the mid-1990s the best
estimates of the ages of the oldest GCs from main sequence turnoff
magnitudes were $t_{GC} \approx 15-16$ Gyr \cite{Bol95,Van,Cha96}. A
frequently quoted lower limit on the age of GCs was $12$ Gyr \cite{Cha96},
which was then an even more conservative lower limit on
$t_0 = t_{GC} + \Delta t_{GC}$, where $\Delta t_{GC} \gsim 0.5$ Gyr is
the time from the Big Bang until GC formation. The main uncertainty in
the GC age estimates came from the uncertain distance to the GCs: a
0.25 magnitude error in the distance modulus translates to a 22\%
error in the derived cluster age \cite{Cha95}.

In spring of 1997, analyses of data from the Hipparcos astrometric
satellite indicated that the distances to GCs assumed in obtaining the
ages just discussed were systematically underestimated
\cite{Reid,Gra}.  It follows that their stars at the main sequence
turnoff are brighter and therefore younger.  Stellar evolution
calculation improvements also lowered the GC age estimates.  In light
of the new Hipparcos data, Chaboyer et al. \cite{Cha98} have done a
revised Monte Carlo analysis of the effects of varying various uncertain
parameters, and obtained $t_{GC} = 11.5 \pm 1.3$ Gyr ($1\sigma$), with
a 95\% C.L.  lower limit of 9.5 Gyr.  The latest detailed analysis
\cite{Caretta} gives $t_{GC} = 11.5\pm2.6$ Gyr from main sequence
fitting using parallaxes of local subdwarfs, the method used in
\cite{Reid,Gra}. These authors get somewhat smaller GC distances when
all the available data is used, with a resulting $t_{GC} = 12.9 \pm
2.9$ Gyr (95\% C.L.).  However, if main sequence fitting is the more
reliable method, the younger age may be more appropriate.

Stellar age estimates are of course based on stellar evolution
calculations, which have also improved significantly.  But the solar
neutrino problem reminds us that we are not really sure that we
understand how even our nearest star operates; and the sun plays an
important role in calibrating stellar evolution, since it is the only
star whose age we know independently (from radioactive dating of early
solar system material).  An important check on stellar ages can come
from observations of white dwarfs in globular and open clusters
\cite{Ric}.

What if the GC age estimates are wrong for some unknown reason? The
only other non-cosmological estimates of the age of the universe come
from nuclear cosmochronometry --- radioactive decay and chemical
evolution of the Galaxy --- and white dwarf cooling. Cosmochronometry
age estimates are sensitive to a number of uncertain issues such as
the formation history of the disk and its stars, and possible actinide
destruction in stars \cite{Mal,Mat}. 
However, an independent cosmochronometry age estimate
of $15.6\pm4.6$ Gyr has been obtained based on data from two
low-metallicity stars, using the measured radioactive depletion of
thorium (whose half-life is 14.2 Gyr) compared to stable heavy
r-process elements \cite{Cow97,Cow99}.  This method could
become very important if it were possible to obtain accurate
measurements of r-process element abundances for a number of very low
metallicity stars giving consistent age estimates, and especially if
the large errors could be reduced.

Independent age estimates come from the cooling of white dwarfs in the
neighborhood of the sun. The key observation is that there is a lower
limit to the luminosity, and therefore also the temperature, of nearby
white dwarfs; although dimmer ones could have been seen, none have
been found (cf. however \cite{Har}).  The only plausible
explanation is that the white dwarfs have not had sufficient time to
cool to lower temperatures, which initially led to an estimate of
$9.3\pm2$ Gyr for the age of the Galactic disk \cite{Win}.
Since there was evidence, based on the pre-Hipparcos GC distances, 
that the stellar disk of our Galaxy is about 2 Gyr younger than the
oldest GCs (e.g., \cite{Ste,Ros}), 
this in turn gave an estimate of the age of the universe of
$t_0 \approx 11\pm2$ Gyr. Other analyses \cite{Wood,Her}
conclude that sensitivity to disk star formation history,
and to effects on the white dwarf cooling rates due to C/O separation
at crystallization and possible presence of trace elements such as
$^{22}$Ne, allow a rather wide range of ages for the disk of about
$10\pm4$ Gyr. One determination of the white dwarf luminosity
function, using white dwarfs in proper motion binaries, leads to a
somewhat lower minimum luminosity and therefore a somewhat higher
estimate of the age of the disk of $\sim 10.5^{+2.5}_{-1.5}$ Gyr
\cite{Osw}.  More recent observations \cite{Leg} and analyses
\cite{Benv} lead to an estimated age of the galactic disk of $8 \pm 1.5$
Gyr.

We conclude that $t_0 \approx 13$ Gyr, with $\sim 10$ Gyr a likely
lower limit. Note that $t_0 > 13$ Gyr implies that $h \leq 0.50$ for
matter density $\Omega_m=1$, and that $h \leq 0.73$ even for $\Omega_m
$ as small as 0.3 in flat cosmologies (i.e., with $\Omega_m +
\Omega_\Lambda = 1$). 

\section{Hubble Parameter $H_0$}

The Hubble parameter $H_0\equiv 100 h$ km s$^{-1}$ Mpc$^{-1}$ remains
uncertain, although no longer by the traditional factor of two.  The
range of $h$ determinations has been shrinking with time \cite{Ken95}.
De~Vaucouleurs long contended that $h \approx 1$. Sandage has long
contended that $h \approx 0.5$, although a recent reanalysis of the
Type Ia supernovae (SNe Ia) data coauthored by Sandage and Tammann
\cite{Saha} concludes that the latest data are consistent with
$h=0.6\pm0.04$.

The Hubble parameter has been measured in two basic ways: (1)
Measuring the distance to some nearby galaxies, typically by measuring
the periods and luminosities of Cepheid variables in them; and then
using these ``calibrator galaxies'' to set the zero point in any of
the several methods of measuring the relative distances to
galaxies. (2) Using fundamental physics to measure the distance to
some distant object(s) directly, thereby avoiding at least some of the
uncertainties of the cosmic distance ladder \cite{Row}.  The
difficulty with method (1) was that there was only a handful of
calibrator galaxies close enough for Cepheids to be resolved in
them. However, the HST Key Project on the Extragalactic Distance Scale
has significantly increased the set of calibrator galaxies.  The
difficulty with method (2) is that in every case studied so far, some
aspect of the observed system or the underlying physics remains
somewhat uncertain. It is nevertheless remarkable that the results of
several different methods of type (2) are rather similar, and indeed
not very far from those of method (1). This gives reason to hope for
convergence.

\subsection{Relative Distance Methods}

One piece of good news is that the several methods of measuring the
relative distances to galaxies now mostly seem to be consistent with
each other. These methods use either ``standard candles'' or empirical
relations between two measurable properties of a galaxy, one
distance-independent and the other distance-dependent. The favorite
standard candle is SNe Ia, and observers are now in good agreement.
Taking account of an empirical relationship between the SNe Ia light
curve shape and maximum luminosity leads to $h = 0.65\pm0.06$
\cite{Rie96}, $h=0.64^{+0.08}_{-0.06}$ \cite{Jha}, or $h =
0.63\pm0.03$ \cite{Ham,Phi}, and the slightly lower value mentioned
above from the latest analysis coauthored by Sandage and Tammann
agrees within the errors.  The HST Key Project result using SNe Ia is
$h = 0.65 \pm 0.02 \pm 0.05$, where the first error quoted is
statistical and the second is systematic \cite{Gib}, and their Cepheid
metallicity-dependent luminosity-period relationship \cite{Ken98} has
been used (this lowers $h$ by 4\%).  Some of the other relative
distance methods are based on old stellar populations: the tip of the
red giant branch (TRGB), the planetary nebula luminosity function
(PNLF), the globular cluster luminosity function (GCLF), and the
surface brightness fluctuation method (SBF). The HST Key Project
result using these old star standard candles is \cite{Fer} $h=0.66\pm
0.04 \pm 0.06$, including the Cepheid metallicity correction. The old
favorite empirical relation used as a relative distance indicator is
the Tully-Fisher relation between the rotation velocity and luminosity
of spiral galaxies.  The ``final'' value of the Hubble constant from
the HST Key Project taking all of these into account, including the
metallicity dependence of the Cepheid period-luminosity relation, is
\cite{Freedman} $h=0.74\pm0.04\pm0.07$, where the first error is
statistical and the second is systematic.  The largest source of
systematic uncertainty is the distance to the LMC, which is here
assumed to have a distance modulus of 18.45.  This is a significantly
higher $h$ than their previous \cite{Mould} $h=0.71\pm0.06$, or
$h=0.68\pm0.06$ including the Cepheid metallicity dependence, using a
LMC distance modulus of 18.5.

\subsection{Fundamental Physics Approaches}

The fundamental physics approaches involve either Type Ia or Type II
supernovae, the Sunyaev-Zel'dovich (S-Z) effect, or gravitational
lensing of quasars.  All are promising, but in each case the relevant
physics remains somewhat uncertain.

The $^{56}$Ni radioactivity method for determining $H_0$ using Type Ia
SNe avoids the uncertainties of the distance ladder by calculating the
absolute luminosity of Type Ia supernovae from first principles using
plausible but as yet unproved physical models for $^{56}$Ni
production.  The first result obtained was that $h=0.61\pm0.10$
\cite{Arn,Bra}; however, another study \cite{Lei} (cf. \cite{Vau}) 
found that
uncertainties in extinction (i.e., light absorption) toward each
supernova increases the range of allowed $h$. Demanding that the
$^{56}$Ni radioactivity method agree with an expanding photosphere
approach leads to $h=0.60^{+0.14}_{-0.11}$ \cite{Nug}. The
expanding photosphere method compares the expansion rate of the SN
envelope measured by redshift with its size increase inferred from its
temperature and magnitude.  This approach was first applied to Type II
SNe; the 1992 result $h=0.6\pm0.1$ \cite{Schmidt}
was subsequently revised upward by the same authors to
$h=0.73\pm0.06\pm0.07$ \cite{Sch94}. However, there are various
complications with the physics of the expanding envelope
\cite{Rui,Eas}.

The S-Z effect is the Compton scattering of microwave background
photons from the hot electrons in a foreground galaxy cluster.  This
can be used to measure $H_0$ since properties of the cluster gas
measured via the S-Z effect and from X-ray observations have different
dependences on $H_0$.  The result from the first cluster for which
sufficiently detailed data was available, A665 (at $z=0.182$), was
$h=(0.4-0.5)\pm0.12$ \cite{Bir91}; combining
this with data on A2218 ($z=0.171$) raised this somewhat to
$h=0.55\pm0.17$ \cite{Bir94}.  The history and more
recent data have been reviewed by Birkinshaw \cite{Bir99}, who concludes
that the available data give a Hubble parameter $h\approx0.6$ with a
scatter of about 0.2.  But since the available measurements are not
independent, it does not follow that $h=0.6\pm0.1$; for example, there
is a selection effect that biases low the $h$ determined this way.

Several quasars have been observed to have multiple images separated
by $\theta \sim$ a few arc seconds; this phenomenon is interpreted as
arising from gravitational lensing of the source quasar by a galaxy
along the line of sight (first suggested by \cite{Ref}; reviewed in
\cite{Wil}).  In the first such system discovered, QSO
0957+561 ($z=1.41$), the time delay $\Delta t$ between arrival at the
earth of variations in the quasar's luminosity in the two images has
been measured to be, e.g., $409\pm23$ days \cite{Pelt},
although other authors found a value of $540\pm12$ days \cite{Pre}.
The shorter $\Delta t$ has now been confirmed \cite{Kun97a,Ser}.
 Since $\Delta t \approx \theta^2 H_0^{-1}$, this
observation allows an estimate of the Hubble parameter. The latest
results for $h$ from 0957+561, using all available data, are $h=0.64
\pm 0.13$ (95\% C.L.) \cite{Kun97a}, and $h=0.62\pm0.07$
\cite{Fal97}, where the error does not include systematic
errors in the assumed form of the lensing mass distribution.

The first quadruple-image quasar system discovered was PG1115+080.
Using a recent series of observations (Schechter et al. 1997), the
time delay between images B and C has been determined to be about
$24\pm3$ days.  A simple model for the lensing galaxy and the nearby
galaxies then leads to $h=0.42\pm0.06$ (Schechter et al. 1997), although
higher values for $h$ are obtained by more sophisticated analyses:
$h=0.60\pm0.17$ \cite{Kee}, $h=0.52\pm0.14$ \cite{Kun97b}.
The results depend on how the lensing galaxy and those in
the compact group of which it is a part are modelled. 

Another quadruple-lens system, B1606+656, leads to $h=0.59 \pm 0.08
\pm 0.15$, where the first error is the 95\% C.L. statistical error,
and the second is the estimated systematic uncertainty \cite{Fas}.
Time delays have also recently been determined for the
Einstein ring system B0218+357, giving $h=0.69^{+0.13}_{-0.19}$ (95\%
C.L.) \cite{Big}.

Mainly because of the systematic uncertainties in modelling the mass
distribution in the lensing systems, the uncertainty in the $h$
determination by gravitational lens time delays remains rather large.
But it is reassuring that this completely independent method gives
results consistent with the other determinations.

\subsection{Conclusions on $H_0$}

To summarize, relative distance methods favor a value $h\approx
0.6-0.8$. Meanwhile the fundamental physics methods typically lead to
$h \approx 0.4-0.7$.  Among fundamental physics approaches, there has
been important recent progress in measuring $h$ via the
Sunyev-Zel'dovich effect and time delays between different images of
gravitationally lensed quasars, although the uncertainties remain
larger than via relative distance methods.  For the rest of this
review, we will adopt a value of $h=0.65\pm0.08$.  This corresponds to
$t_0= 6.52 h^{-1} {\rm Gyr} = 10 \pm 2$ Gyr for $\Omega_m=1$ ---
probably too low compared to the ages of the oldest globular clusters.
But for $\Omega_m=0.2$ and $\Omega_\Lambda=0$, or alternatively for
$\Omega_m=0.4$ and $\Omega_\Lambda=0.6$, $t_0 = 13\pm2$ Gyr, in
agreement with the globular cluster estimate of $t_0$.  This is one of
the weakest of the several arguments for low $\Omega_m$, a non-zero
cosmological constant, or both.

\section{Hot Dark Matter Density $\Omega_\nu$}

The recent atmospheric neutrino data from Super-Kamiokande
\cite{Fukuda98} provide strong evidence of neutrino oscillations and
therefore of non-zero neutrino mass.  These data imply a lower limit
on the hot dark matter (i.e., light neutrino) contribution to the
cosmological density $\Omega_\nu \gsim 0.001$.  $\Omega_\nu$ is
actually that low, and therefore cosmologically uninteresting, if
$m(\nu_\tau) \gg m(\nu_\mu)$, as is suggested by the hierarchical
pattern of the quark and charged lepton masses.  But if the $\nu_\tau$
and $\nu_\mu$ are nearly degenerate in mass, as suggested by their
strong mixing, then $\Omega_\nu$ could be substantially larger.
Although the Cold + Hot Dark Matter (CHDM) cosmological model with
$h\approx 0.5$, $\Omega_m=1$, and $\Omega_\nu=0.2$ predicts power
spectra of cosmic density and CMB anisotropies that are in excellent
agreement with the data \cite{P97,Gaw}, as we have just seen the large
value measured for the Hubble parameter makes such $\Omega_m=1$ models
dubious.  It remains to be seen whether including a significant amount
of hot dark matter in low-$\Omega_m$ models improves their agreement
with data.  Primack \& Gross \cite{P98,PG00} found that the possible
improvement of the low-$\Omega_m$ flat (\lcdm) cosmological models
with the addition of light neutrinos appears to be rather limited, and
the maximum amount of hot dark matter decreases with decreasing
$\Omega_m$ \cite{P95}.  For $\Omega_m \lsim 0.4$, \cite{Cro} find that
$\Omega_\nu \lsim 0.08$; \cite{Fuk} finds more restrictive upper
limits with the constraint that the primordial power spectrum index $n
\le 1$, but this may not be well motivated.

\section{Cosmological Constant $\Lambda$}

The strongest evidence for a positive $\Lambda$ comes from
high-redshift SNe Ia, and independently from a combination of
observations indicating that $\Omega_m \sim 0.4$ together with CMB
data indicating that the universe is nearly flat.  We will discuss
these observations in the next section.  Here we will start by looking
at other constraints on $\Lambda$.

The cosmological effects of a cosmological constant are not difficult
to understand \cite{Fel,Lah,Carroll}.
In the early universe, the density of energy
and matter is far more important than the $\Lambda$ term on the
r.h.s. of the Friedmann equation. But the average matter density
decreases as the universe expands, and at a rather low redshift ($z
\sim 0.2$ for $\Omega_m=0.3$, $\Omega_\Lambda=0.7$) the $\Lambda$ term
finally becomes dominant.  Around this redshift, the $\Lambda$ term
almost balances the attraction of the matter, and the scale factor $a
\equiv (1+z)^{-1}$ increases very slowly, although it ultimately
starts increasing exponentially as the universe starts inflating under
the influence of the increasingly dominant $\Lambda$ term.  The
existence of a period during which expansion slows while the clock
runs explains why $t_0$ can be greater than for $\Lambda=0$, but this
also shows that there is an increased likelihood of finding galaxies
in the redshift interval when the expansion slowed, and a
correspondingly increased opportunity for lensing by these galaxies of
quasars (which mostly lie at higher redshift $z \gsim 2$).

The observed frequency of such optical lensed quasars is about what
would be expected in a standard $\Omega=1$, $\Lambda=0$ cosmology, so
this data sets fairly stringent upper limits: $\Omega_\Lambda \leq
0.70$ at 90\% C.L. \cite{Maoz,Koc93}, with more recent data giving
even tighter constraints: $\Omega_\Lambda < 0.66$ at 95\% confidence
if $\Omega_m + \Omega_\Lambda =1$ \cite{Koc96}.  This limit could
perhaps be weakened if there were (a) significant extinction by dust
in the E/S0 galaxies responsible for the lensing or (b) rapid
evolution of these galaxies, but there is much evidence that these
galaxies have little dust and have evolved only passively for $z \lsim
1$ \cite{Steidel,Lil,Schade}.  An alternative analysis \cite{Im} of
some of the same optical lensing data gives a value $\Omega_\Lambda
=0.64_{-0.26}^{+0.15}$.  My group \cite{Maller} (cf. \cite{BartLoeb})
showed that edge-on disk galaxies can lens quasars very effectively,
and discussed a case in which optical extinction is significant.  But
the radio observations discussed by \cite{Fal97}, which give a
$2\sigma$ limit $\Omega_\Lambda < 0.73$, are not affected by
extinction, so those are the ones quoted in the Table above.  Recently
a reanalysis \cite{Chi} of lensing using new models of the evolution
of elliptical galaxies gave $\Omega_\Lambda=0.7^{+0.1}_{-0.2}$, but
Kochanek et al. \cite{Koc99} (see especially Fig. 4) show that the
available evidence disfavors such models.

A model-dependent constraint appeared to come from simulations of
$\Lambda$CDM \cite{KPH} and OpenCDM \cite{Jen}
COBE-normalized models with $h=0.7$, $\Omega_m=0.3$, and
either $\Omega_\Lambda=0.7$ or, for the open case, $\Omega_\Lambda=0$.
These models have too much power on small scales to be consistent with
observations, unless there is strong scale-dependent antibiasing of
galaxies with respect to dark matter.  However, recent high-resolution
simulations \cite{KGKK} find that merging and destruction of
galaxies in dense environments lead to exactly the sort of
scale-dependent antibiasing needed for agreement with observations for
the \lcdm\ model. Similar results have been found using simulations
plus semi-analytic methods \cite{Benson} (but cf. \cite{Kau}).

Another constraint on $\Lambda$ from simulations is a claim \cite{Bar}
that the number of long arcs in clusters is in accord with
observations for an open CDM model with $\Omega_m=0.3$ but an order of
magnitude too low in a \lcdm\ model with the same $\Omega_m$.  This
apparently occurs because clusters with dense cores form too late in
such models.  This is potentially a powerful constraint, and needs to
be checked and understood.  It is now known that including cluster
galaxies does not alter these results \cite{Men,Flo}.

\section{Measuring $\Omega_m$}

The present author, like many theorists, has long regarded the
Einstein-de Sitter ($\Omega_m=1$, $\Lambda=0$) cosmology as the most
attractive one. For one thing, of the three possible constant
values for $\Omega$ --- 0, 1, and $\infty$ --- the only one
that can describe our universe is $\Omega_m=1$.  Also, cosmic
inflation is the only known solution for several otherwise intractable
problems, and all simple inflationary models predict that the universe
is flat, i.e. that $\Omega_m + \Omega_\Lambda=1$.  Since there is no
known physical reason for a non-zero cosmological constant, it was
often said that inflation favors $\Omega=1$.  Of course, theoretical
prejudice is not a reliable guide.  In recent years, many cosmologists
have favored $\Omega_m \sim 0.3$, both because of the $H_0-t_0$
constraints and because cluster and other relatively small-scale
measurements have given low values for $\Omega_m$. (For a summary of
arguments favoring low $\Omega_m \approx 0.2$ and $\Lambda=0$, see
\cite{Coles}; \cite{Dek97} is a review that notes that larger scale
measurements favor higher $\Omega_m$.)
But the most exciting new evidence has come from
cosmological-scale measurements.

{\bf Type Ia Supernovae.}  At present, the most promising techniques
for measuring $\Omega_m$ and $\Omega_\Lambda$ on cosmological scales
use the small-angle anisotropies in the CMB radiation and
high-redshift Type Ia supernovae (SNe Ia).  We will discuss the latter
first.  SNe Ia are the brightest supernovae, and the spread in their
intrinsic brightness appears to be relatively small.  The Supernova
Cosmology Project \cite{Per97a} demonstrated the feasibility of
finding significant numbers of such supernovae.  The first seven high
redshift SNe Ia that they analyzed gave for a flat universe
$\Omega_m=1- \Omega_\Lambda= 0.94^{+0.34}_{-0.28}$, or equivalently
$\Omega_\Lambda= 0.06^{+0.28}_{-0.34}$ ($<0.51$ at the 95\% confidence
level) \cite{Per97a}.  But adding one $z=0.83$ SN Ia for which they
had good HST data lowered the implied $\Omega_m$ to $0.6\pm0.2$ in the
flat case \cite{Per97b}.  Analysis of their larger dataset of 42
high-redshift SNe Ia gives for the flat case $\Omega_m = 0.28^{+0.09
+0.05}_{-0.08 -0.04}$ where the first errors are statistical and the
second are identified systematics \cite{Per99}.  The High-Z Supernova
team has also searched successfully for high-redshift supernovae to
measure $\Omega_m$ \cite{Gar,Rie98}, and their 1998 dataset of $14+2$
high-redshift SNe Ia including three for which they had HST data (two
at $z\approx 0.5$ and one at 0.97) imply $\Omega_m=0.32 \pm 0.1$ in
the flat case with their MLCS fitting method.

The main concerns about the interpretation of this data are evolution
of the SNe Ia \cite{Dre,Rierev} and dimming by dust.  A recent
specific supernova evolution concern is that the rest frame rise-times
of distant supernovae may be longer than nearby ones \cite{Rie99}.
But a direct comparison between nearby supernova and the SCP distant
sample shows that they are rather consistent with each other
\cite{Ald}.  Ordinary dust causes reddening, but hypothetical ``grey''
dust would cause much less reddening and could in principle provide an
alternative explanation for the fact that high-redshift supernovae are
observed to be dimmer than expected in a critical-density cosmology.
Grey interstellar dust would induce more dispersion than is observed,
so the hypothetical grey dust would have to be intergalactic.  It is
hard to see why the largest dust grains, which would be greyer, should
preferentially be ejected by galaxies \cite{Sim}.  Such dust, if it
exists, would also absorb starlight and reradiate it at long
wavelengths, where there are other constraints that could, with
additional observations, rule out this scenario \cite{Agu}.  Such grey
dust would also produce some reddening which could be detectable via
comparison of infrared vs. optical colors of supernovae; such a
measurement for one high-redshift SN Ia disfavors significant grey
dust extinction \cite{Riedust}, and more observations could strengthen
this conclusion.  Yet another way of addressing this question is to
collect data on supernovae with redshift $z>1$, where the dust
scenario predicts considerably more dimming than the $\Lambda$
cosmology.  The one $z>1$ supernova currently available, SCP's
``Albinoni'' (SN1998eq) at $z=1.2$, favors the $\Lambda$ cosmology.
More such data are needed for a statistically significant result, and
both the SCP and the High-Z group are attempting to get a few more
very high redshift supernovae.

{\bf CMB anisotropies.}  The location of the first acoustic (or
Doppler, or Sak\-ha\-rov) peak at angular wavenumber $l\approx 200$
indicated by the data available at the time of this meeting was
evidence in favor of a flat universe $\Omega_{tot} \equiv \Omega_m +
\Omega_\Lambda \approx 1$ (e.g. \cite{Dod}).  New data from the
BOOMERANG long-duration balloon flight around Antarctica \cite{boom}
and the MAXIMA-1 balloon flight \cite{maxima} confirm this, with
$\Omega_{tot} = 1.11^{+0.13}_{-0.12}$ at 95\% C.L. \cite{jaffe}.  The
preliminary BOOMERANG results \cite{boom} are lower around $l \approx
500$ than the predictions in this second peak region in \lcdm-type
models (e.g., \cite{Hu}), and this could \cite{boomparam} indicate
higher $\Omega_b$ than expected from Big Bang Nucleosynthesis together
with the recent deuterium measurements (discussed below).  However,
the MAXIMA-1 data for $l \approx 500$ are more consistent with
expectations of standard models and the standard BBN $\Omega_b$
\cite{maximaparam} (but cf. \cite{jaffe}).  
The BOOMERANG and MAXIMA-2 data are still being
analyzed, and other experiments will have relevant data as well.
Further data should be available in 2001 from the NASA Microwave
Anisotropy Probe satellite.

{\bf Large-scale Measurements.} The comparison of the IRAS redshift
surveys with POTENT and related analyses typically give values for the
parameter $\beta_I \equiv \Omega_m^{0.6}/b_I$ (where $b_I$ is the
biasing parameter for IRAS galaxies), corresponding to $0.3 \lsim
\Omega_m \lsim 3$ (for an assumed $b_I=1.15$).  It is not clear
whether it will be possible to reduce the spread in these values
significantly in the near future --- probably both additional data and
a better understanding of systematic and statistical effects will be
required.  A particularly simple way to deduce a lower limit on
$\Omega_m$ from the POTENT peculiar velocity data was proposed by
\cite{Dek94}, based on the fact that high-velocity outflows from voids
are not expected in low-$\Omega$ models.  Data on just one nearby void
indicates that $\Omega_m \ge 0.3$ at the 97\% C.L.  Stronger
constraints are available if we assume that the probability
distribution function (PDF) of the primordial fluctuations was
Gaussian.  Evolution from a Gaussian initial PDF to the non-Gaussian
mass distribution observed today requires considerable gravitational
nonlinearity, i.e. large $\Omega_m$.  The PDF deduced by POTENT from
observed velocities (i.e., the PDF of the mass, if the POTENT
reconstruction is reliable) is far from Gaussian today, with a long
positive-fluctuation tail.  It agrees with a Gaussian initial PDF if
and only if $\Omega_m \sim 1$; $\Omega_m <1$ is rejected at the
$2\sigma$ level, and $\Omega_m \leq 0.3$ is ruled out at $\ge 4\sigma$
\cite{Nus,Ber}.  It would be interesting to repeat this analysis with
newer data.  Analyzing peculiar velocity data without POTENT again
leads to a strong lower limit $\Omega_m > 0.3$ (99\% C.L.), and together
with the SN Ia constraints leads to the conclusion that $\Omega_m
\approx 0.5$ \cite{ZehaviD}.

{\bf Measurements on Scales of a Few Mpc.} A study by the Canadian
Network for Observational Cosmology (CNOC) of 16 clusters at $z\sim
0.3$, mostly chosen from the Einstein Medium Sensitivity Survey 
\cite{Hen}, was designed to allow a self-contained measurement of
$\Omega_m$ from a field $M/L$ which in turn was deduced from their
measured cluster $M/L$. The result was $\Omega_m=0.19\pm0.06$
\cite{Carlberg}.  These data were mainly compared to
standard CDM models, and they appear to exclude $\Omega_m=1$ in such
models.

{\bf Estimates on Galaxy Halo Scales.} Work by Zaritsky et al. \cite{Zar93}
has confirmed that spiral galaxies have massive halos.  They collected
data on satellites of isolated spiral galaxies, and concluded that the
fact that the relative velocities do not fall off out to a separation
of at least 200 kpc shows that massive halos are the norm.  The
typical rotation velocity of $\sim 200-250$ km s$^{-1}$ implies a mass
within 200 kpc of $\sim 2\times10^{12} M_\odot$.  A careful analysis
taking into account selection effects and satellite orbit
uncertainties concluded that the indicated value of $\Omega_m$ exceeds
0.13 at 90\% confidence \cite{Zar94}, with preferred
values exceeding 0.3. Newer data suggesting that relative velocities
do not fall off out to a separation of $\sim 400$ kpc \cite{Zar97}
presumably would raise these $\Omega_m$ estimates.  Weak lensing
data confirms the existence of massive galactic
halos \cite{sdsswklens,vanw,bacon,whit}.

{\bf Cluster Baryons vs. Big Bang Nucleosynthesis.}  White et al.
\cite{Whi} emphasized that X-ray observations of the abundance of baryons
in clusters can be used to determine $\Omega_m$ if clusters are a fair
sample of both baryons and dark matter, as they are expected to be
based on simulations \cite{Evr}. The fair sample hypothesis implies that
\begin{equation}
\Omega_m = \frac{\Omega_b}{f_b} = 0.3 \left(\frac{\Omega_b}{0.04}\right)
                                  \left(\frac{0.13}{f_b}\right).
\end{equation}
We can use this to determine $\Omega_m$ using the baryon abundance
$\Omega_b h^2 = 0.019 \pm 0.0024$ (95\% C.L.) from the measurement of
the deuterium abundance in high-redshift Lyman limit systems, of which
a third has recently been analyzed \cite{Kir,Tytler00} and more are in
the pipeline {D. Tytler, these proceedings}.  Using X-ray data from an
X-ray flux limited sample of clusters to estimate the baryon fraction
$f_b = 0.075 h^{-3/2}$ \cite{Mohr} gives $\Omega_m = 0.25 h^{-1/2} =
0.3\pm0.1$ using $h=0.65\pm0.08$.  Estimating the baryon fraction
using Sunyaev-Zel'dovich measurements of a sample of 18 clusters gives
$f_b = 0.077 h^{-1}$ \cite{Carlstrom}, and implies $\Omega_m = 0.25
h^{-1} = 0.38\pm0.1$.

{\bf Cluster Evolution.}  The dependence of the number of clusters on
redshift can be a useful constraint on theories \cite{Eke96}.
But the cluster data at various redshifts are difficult to
compare properly since they are rather inhomogeneous.  Using just
X-ray temperature data, \cite{Eke98} concludes that $\Omega_m
\approx 0.45\pm0.2$, with $\Omega_m=1$ strongly disfavored.

{\bf Power Spectrum.} In the context of the \lcdm\ class of models,
two additional constraints are available.  The spectrum shape
parameter $\Gamma \approx \Omega_m h \approx 0.25\pm0.05$, implying
$\Omega_m \approx 0.4\pm0.1$.  A new measurement $\Omega_m = 0.34\pm0.1$
comes from the amplitude of the power spectrum of fluctuations at
redshift $z\sim3$, measured from the Lyman $\alpha$ forest \cite{Wei}.
This result is strongly inconsistent with
high-$\Omega_m$ models because they would predict that the
fluctuations grow much more to $z=0$, and thus would be lower at $z=3$
than they are observed to be.

\begin{figure}
\centering
\centerline{\psfig{file=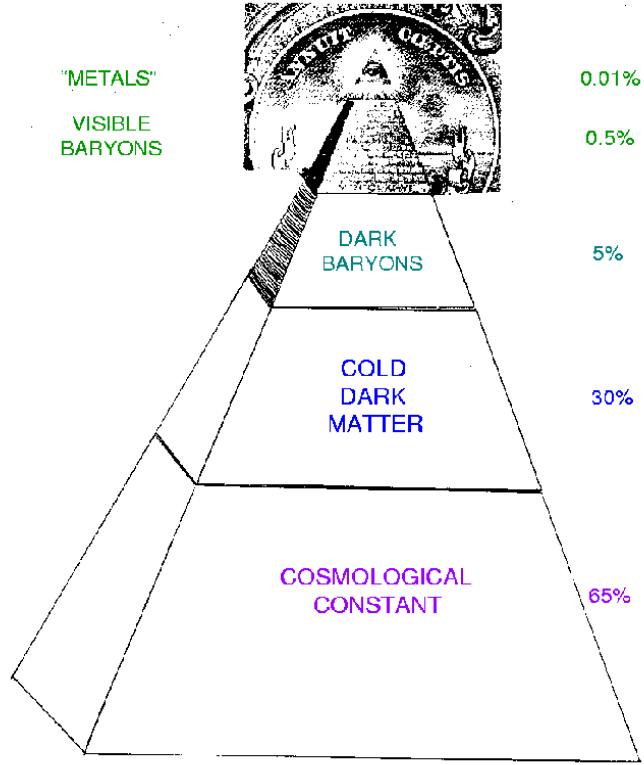,height=11cm}}
\caption{ 
The Great Seal of the United States, found on the back of the American
dollar bill, includes a pyramid representing strength and duration,
capped by the eye of Providence.  Here we use this to represent the
visible matter in the universe ($\Omega_{vis} \approx 0.005$), with the
upper triangle containing the eye representing the metals (elements
heavier than hydrogen and helium, with $\Omega_{metals}\approx
10^{-4}$) since most of the mass of our bodies is made up of these 
elements.  The three-dimensional nature of the pyramid, which here
continues below the part shown on the Great Seal, makes it useful for
showing graphically the relative proportions of the dark baryons, cold
dark matter, and cosmological constant (or dark energy).
}
\end{figure}

\section{Conclusion}

We thus end up with a picture of the distribution of the density of
energy density in a flat universe represented by Figure 1
\cite{CosQuestions}.  One of the most striking things about the
present era in cosmology is the remarkable agreement between the
values of the cosmological densities and the other cosmological
parameters obtained by different methods --- except possibly for the
quasar lensing data which favors a higher $\Omega_m$ and lower
$\Omega_\Lambda$, and the arc lensing data which favors lower values
of both parameters.  If the results from the new CMB measurements end
up agreeing with those from the other methods discussed above, the
cosmological parameters will have been determined to perhaps 10\%, and
cosmologists can focus their attention on the other subjects that I
mentioned at the beginning: origin of the initial fluctuations, the
nature of the dark matter and dark energy, and the formation of
galaxies and large-scale structure.  Cosmologists can also speculate
on the reasons why the cosmological parameters have the values that
they do, but this appears to be the sort of question whose answer may
require a deeper understanding of fundamental physics --- perhaps from
a superstring theory of everything.

%\acknowledgments 
This work was supported in part by NSF and NASA grants
and a faculty grant at UCSC.  I am grateful to Leo Stodolsky for 
hospitality at the Max Planck Institute for Physics in Munich, and
to the Alexander von Humboldt Foundation for a Humboldt Award.

\end{document}